# Color Blending in Outdoor Optical See-through AR: The Effect of Real-world Backgrounds on User Interface Color

Joseph L. Gabbard, Member IEEE, J. Edward Swan II, Senior Member, IEEE, and Adam Zarger

**Abstract**— It has been noted anecdotally and through a small number of formal studies that ambient lighting conditions and dynamic real-world backgrounds affect the usability of optical see-through augmented reality (AR) displays; especially so in outdoor environments. Our previous work examined these effects using painted posters as representative real-world backgrounds [1]. In this paper, we present a study that employs an experimental testbed that allows AR graphics to be overlaid onto real-world backgrounds as well as painted posters. Our results indicate that color blending effects of physical materials as backgrounds are nearly the same as their corresponding poster backgrounds, even though the colors of each pair are only a metameric match. More importantly, our results suggest that given the current capabilities of optical see-through head-mounted displays (oHMDs), the implications are, at a minimum, a reduced color gamut available to user interface (UI) designers. In worse cases, there are unknown or unexpected color interactions that no UI or system designers can plan for; significantly crippling the usability of the UI or altering the semantic interpretation of graphical elements. Further, our results support the concept of an adaptive AR system which can dynamically alter the color of UI elements based on predicted background color interactions. These interactions can be studied and predicted through methods such as those presented in this work.

✦

## 1 INTRODUCTION

During the past 20 or so years, AR technologies have slowly made their way into several application domains (e.g., Okur et al. [2]; Debenham, Thomas & Trout [3]; Livingston et al. [4], Gleue and Dähne [5], Caudell and Mizell [6]). More recently, AR concepts and applications have begun to gain renewed momentum in the public eye; be it through the proliferation of location-aware handheld AR apps [7], growing numbers of AR demonstrations on YouTube, television advertisements and live sporting events that continue to leverage AR concepts, penetration of more sophisticated video-AR systems into the mainstream automobile industry (e.g., overlaid graphics onto rear- and side-facing cameras), and most recently, Google's announcement of Project Glass (an optical-see through head-mounted display with voice-activated smartphone, GPS and internet capabilities). As has been seen in earlier AR system and with other technologies (e.g., virtual reality and artificial intelligence), there are well-founded concerns that AR may be over-promised or at least, over-perceived. And while significant progress has been made on technically-limiting hurdles such as outdoor tracking and registration, we must be careful not to ignore other equally challenging and potentially crippling AR research areas as AR attempts to is affirm its position as a viable consumer technology platform.

With significant improvements in mobile phone and tablet displays, user expectations of display performance is at an all-time high. Classical display characteristics such as resolution, clarity, color, contrast and (for head-worn displays) field of view, etc., likely play an important roles in meeting these expectations. Many of these technical features are largely controlled by display design and specification (resolution, fov, etc.). Others, such as color, luminance and thus contrast, manifest themselves differently depending on the context (absolute dark vs. daylight) and significant interactions with the real world. Currently, we do not fully understand how even the most advanced optical see-through display's color and luminance systems interact with environmental lighting and backgrounds that effectively act as the AR systems canvas or "desktop".

Traditional UI designers, are afforded the luxury of being able to carefully control the color each pixel atop a black (blank) canvas in order to carefully craft elegant designs and effective UIs (Figure 1). In video see-through AR, UI designers must take into account dynamic backgrounds and lighting, but in the end are able to completely "overwrite" any given camera-captured real-world pixel and/or augmented overlaid graphics pixel color and transparency to mitigate problems with color perception and/or contrast to ultimately achieve a desired effect. Moreover, in both traditional UIs and video-based AR systems, designers are privy to a rich color-palette, ranging from black (RGB: 0, 0, 0) to white (RGB: 255, 255, 255), and including over 65 million colors. In optical see-through displays, UI designers are at a relative disadvantage. Since their canvas is not black (it is the optical light from real-world), nor is it completely "overwrite-able", even the most saturated pixel color and/or most illuminated pixel rendered via today's oHMD displays result in some level of graphics transparency onto the real-world. Indeed, it is this capability of transparency that makes optical see-through AR applications appealing in many contexts. Aside from the dynamic and, to some extent, uncontrollable canvas (i.e., real-world background), oHMD UI designers also have a much more limited color palette in which to work since the color black is essentially completely transparent using additive color oHMD displays. As such, there is a large set of non-black colors (to date, the actual number has not been quantified) with low luminance (i.e., "dark colors") that in other display paradigms afford good contrast against bright backgrounds, yet are simply unavailable or un-useable. To exacerbate the challenge, the colors that *are* available to the oHMD UI designer may be perceived differently from moment to moment depending upon the constantly changing canvas (real-world background). How can oHMD UI designers create effective and semantically meaningful UI designs without understanding how these designs may be perceived by different users under different conditions? Currently, one strategy is to design for specific usage contexts where the backgrounds are likely known, and may not change that often either in chrominance or luminance, e.g., in an desert environment. Another popular strategy (for oHMD text color, contrast and legibility) is to surround the graphical element in a large white

- *Joseph L. Gabbard is with Virginia Tech.*
  *E-mail: jgabbard@vt.edu.*
- *J. Edward Swan II is with Mississippi State University.*
  *E-mail: swan@acm.org*
- *Adam Zarger is with Excella Consulting and formally with Virginia Tech.*
  *E-mail: adam.zarger@excella.com*

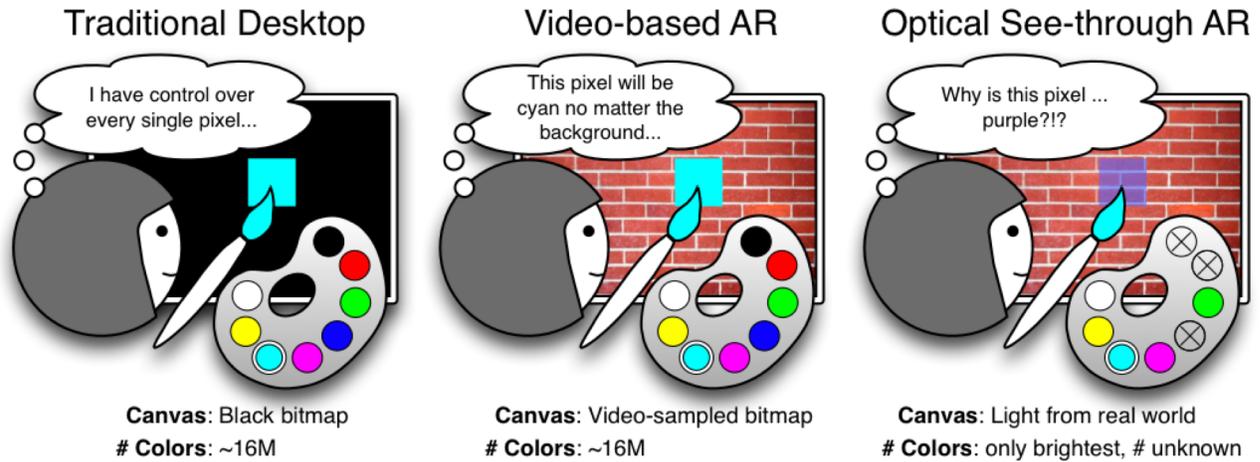

**Figure 1**. The designer's canvas in outdoor optical see-through AR settings is much more challenging that that of traditional or even video-see through AR. Specifically, the color palette is greatly reduced, and there is no certainty that the color used for an interface element will be preserved given dynamic changes in real-world backgrounds and lighting conditions.

billboard; however such strategies occlude more of the real world than may be necessary. Longitudinally, our work aims to better understand these color and luminance interactions, and ultimately create a model capable of taking a UI designers intent (e.g., I want *this* pixel to be perceived as *blue* by users with normal vision) and dynamically rendering it (e.g., using real-time adaptive algorithms) to an oHMD in order to best preserve and maintain that intent.

In this paper, we examine and *quantify*, how environmental and display-generated light combine in an oHMD, given a set of representative (proxy and physical/actual) real-world backgrounds. Section 2 presents related work in AR and other fields. In section 3, we discuss our optical testbed that emulates outdoor lighting conditions and allows us to carefully measure the combined color of virtual colors and real-world backgrounds as projected through an optical see-through display. Section 4 describes our study, including the experimental design, how materials were assembled to create real-world backgrounds, and data collection methods. We present a rich set of results and discussion in Section 5, including our approach to qualitative and quantitative data analysis, followed by ideas for Future Work in Section 6.

## 2 RELATED WORK

While not fully understood, recent work in AR has undoubtedly shown that the blending of real-world lighting, backgrounds, and virtual graphics has an effect on display usability and user performance (Kerr et al. [8], Kalkofen [9], Gabbard et al. [1, 10, 11], Livingston et al. [12], Peterson, Axholt & Ellis [13], Pingel & Clarke [14], Thomas et al. [15]). In many cases, outdoor environmental conditions can, for example, dramatically alter users' color perception of user interface elements. A common observation in a number of studies using outdoor AR using oHMDs is that text or icon colors become washed out in outdoor conditions.

Kerr et al. [8] present a study a employing a monocular oHMD in an outdoor urban environment, where participants complete wayfinding tasks while relying on augmented imagery for orientation. The authors emphatically state that the most important issue they encountered was the inability of participants to see the display clearly outdoors, adding that *all* participants encountered difficulty using the oHMD in bright outdoor environments with some graphical elements more visible than others.

Pingel & Clarke [14] noted that participants had difficulty, measured both quantitatively and qualitatively, locating waypoint markers due to issues with the colors on their AR map display. They further consider mitigating the problem by choosing colors that "stand out" all the time, or by using adaptive cartographic (color) representations, but note users' difficulty to cognitively follow constantly changing symbol colors.

Thomas et al. [15] conducted an informal study to determine a set of initial colors to graphically encode an outdoor AR user interface. Their goal was to assess the opaqueness and visibility of augmenting graphics, under a variety of viewing conditions (shade-to-sun, sun-to-shade, etc.). After examining 36 AR colors, they listed a small number of usable colors for specific viewing conditions, but do not give details on the real-world background on which their results are based (although some rough assumptions can be made by examining their figures and usage content).

Kalkofen et al. [9] addresses similar color issues in mixed-reality, and warns readers that "careless generation of visualization in AR environments may lead to misleading interactions of colors and shades representing the real and the virtual objects." They further assert that renderings of virtual objects that do not take the prospective real world surroundings into account may result in a misinterpretation of intention of the composed virtual and real scene.

These insights speak directly to the issue of color blending with respect to UI color encoding and color perception; specifically the fact that semantic interpretation of interface elements can, and quite often will, be compromised – resulting in severe usability problems or critical incidents in application domains where color encoding is critical (e.g., military, medical visualization, etc.). While numerous researchers have seen this anecdotally, *there is still much work to be done to fully understand this phenomenon*.

Some work has been done to quantify the effects of real-world background and lighting on virtual graphics. Gabbard et al. [1] present an engineering study that uses a colorimeter to measure the light that enters a users eye given an AR display color and a representative real-world background. They measured over two dozen AR display colors against several poster-based backgrounds. Their results show a strong interaction between background and AR color, and more specifically show how the color gamut for oHMDs in outdoor settings can be significantly compressed around the whitepoint.

Livingston et al. [12] also examines the combined effect display contrast and resolution on visual acuity using a varied set of different AR displays. They describe a color matching study that suggests AR users perceptually distort some colors (e.g., blue) more than others (e.g., the physical color swatch a user specifies to match an AR color is much different than the target color). Their user studies described therein further suggest that modest adjustments to levels of contrast can increase acuity in visual tasks.

Peterson, Axholt & Ellis [13] present a novel technique for segregating overlapping labels for far-field objects in outdoor AR. In this work, they note the difficulty in maintaining legible text due to the dynamic outdoor conditions (e.g., passing clouds and movement of the sun during the day), and ultimately vary the number of neutral density filters used to ensure visibility both of the virtual images and the background markers. For their study, it was necessary to keep the virtual display elements at least 10-15% brighter than the background.

Gabbard, Swan, and Hix [11] present an oHMD-based outdoor study that examined the effect of text color and backgrounds on a text legibility task. The study was extended by Gabbard, Swan, et al. [10] to include additional independent variables and actual real-world backgrounds. Results from these studies indicate that backgrounds affect perceived text color and thus legibility in outdoor AR reading tasks.

## 3 A Testbed for Conducting Color Blending Engineering Studies

Our goals for creating the testbed were to (1) simulate natural outdoor lighting *quality* and *brightness* in a controlled indoor environment, (2) systematically vary real-world background objects, and thus systematically vary reflected, colored light, (3) integrate an oHMD display, driven by software capable of systematically displaying virtual stimuli, (4) determine a data collection method that is accurate, reliable, and automated as much as possible, and (5) establish a research testbed flexible enough to conduct both engineering studies, as well as, planned user-based studies.

The testbed is mounted on an optical bench so that we can precisely adjust and control the position and orientation of each testbed component (Figure 2). The testbed supports systematic manipulation of AR light and light reflected off actual physical materials (e.g., real bricks, pavement cut from a road, etc.) as real-world backgrounds, as well as painted posters used as representative or proxy real-world backgrounds.

We used specialized lights that reproduce the daylight standard white point of D65, and accurately project as much of the visible color spectrum as naturally exists outdoors. Light sources are rated using the color rendering index (CRI), which ranges from 1–100, with 100 denoting a light that can perfectly reproduce the entire daylight color spectrum. For this study, we used fluorescent lights rated at 95 CRI. We sealed the lights in an enclosure constructed of black foam core board. We cut a large rectangular hole in the back of the enclosure to allow physical materials to be placed flush up against the back. For a more thorough discussion of the enclosure's lighting and general construction see [1].

For our testbed, we disassembled an NVIS nVisor SX oHMD and custom mounted its monocle inline with other testbed components. The oHMD monocle is designed to be 23 millimeters from a user's eye, where the resulting display projection creates a footprint smaller than what is required by our colorimeter sensor. As such, we placed a 50 mm camera SLR lens at a location approximate to a user's eye when wearing the display to increase the area in which to sample. We mounted an OTC1000 ColorPro Optical Color Analyzer (colorimeter) downstream of the SLR lens to measure the blended light. Both the SLR lens and the colorimeter were mounted on precision optical bench hardware allowing for fine tuning along 5 degrees of freedom (both components were fixed in-line along one positional axis as shown in Figure 2).

Lastly, we developed a calibration process for the testbed to ensure our components were positioned correctly, and that data collection was as accurate and reproducible as possible. We based the calibration process on the D65 white point, verifying white point measurements (and slightly adjusting testbed components as needed) under three different white point measurement conditions: oHMD projecting all white graphics with enclosure lights off; oHMD off with enclosure lights on; and oHMD projecting all white graphics with enclosure lights on. By examining all three conditions, we were assured that data collected under any experimental lighting condition were accurate and consistent with respect to the D65 white point. We further verified the calibration by separately projecting red, green and blue oHMD light under the no-lights (enclosure lights off) condition, checking the three extreme colorimeter measurements against the respective red, green, blue corners of the standard sRGB color gamut (i.e., per oHMD color specification).

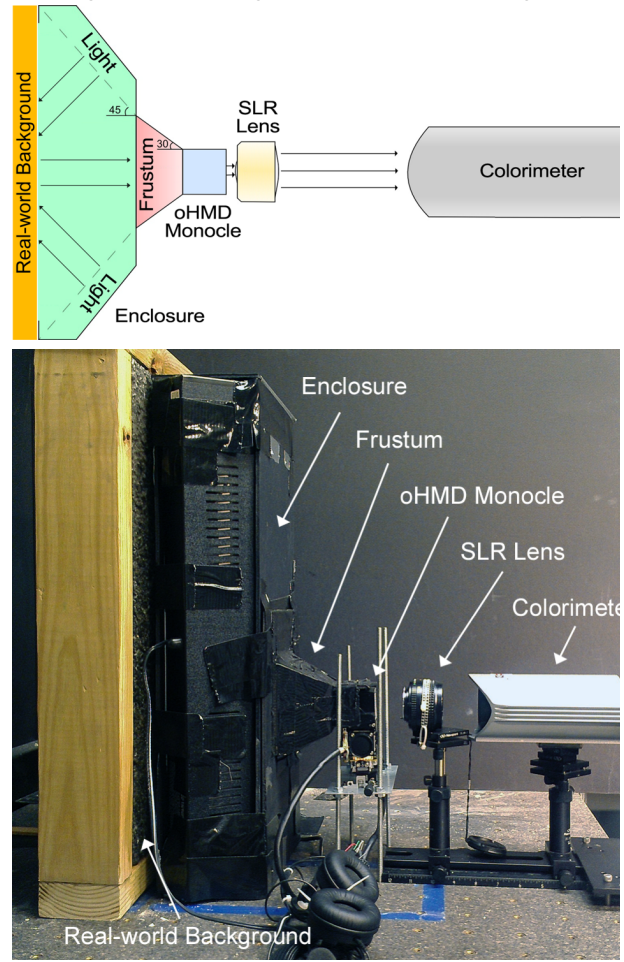

**Figure 2**. (Above) A top view schematic of our experimental testbed depicting an enclosure (green) containing two 15-watt lights reflecting off of real-world backgrounds (orange). Indirectly reflected light exited the enclosure through the frustum (red) and into the AR oHMD monocle (blue). The AR oHMD image, designed to be focused on a user's retina, was directed towards the colorimeter (grey) using an SLR camera lens (yellow). A desktop computer (not shown) was used to drive the AR display, as well as, to capture/log data from the colorimeter. (Below) An annotated photograph of our optical experimental testbed shown with pavement material as real-world background.

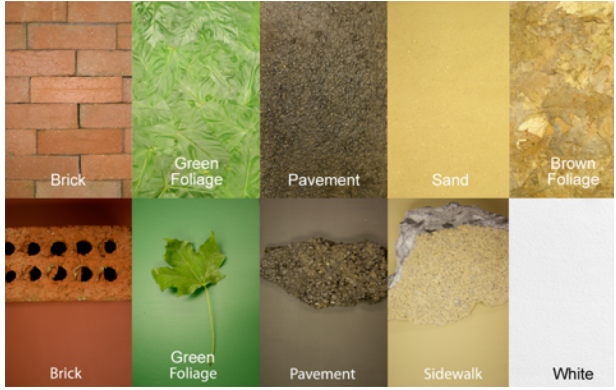

**Figure 3.** We created five real-world backgrounds using physical materials (top row): *brick*, *green foliage*, *pavement*, *sand* and *brown foliage*. These backgrounds were mounted in modular frames so we could easily place them at the back of the enclosure. We also included four painted posters (bottom row): *brick*, *green foliage*, *pavement* and *sidewalk* as additional backgrounds. We used a blank poster for the *white* background condition. The eleventh background was a no-lights condition.

**Figure 4.** We tested 27 colors, defined by every combination of red (R), green (G), and blue (B), at the levels of 0, 128, and 255. For the color black, the oHMD was turned off. Note that the colors shown here, as well as elsewhere in this paper, are just representative of the real-world experience of seeing that color.

| R | G | B | Name | R | G | B | Name |
|---|---|---|---|---|---|---|---|
| off | off | off | black | 0 | 255 | 255 | cyan |
| 128 | 0 | 0 | maroon | 0 | 128 | 255 | azure |
| 128 | 128 | 0 | olive | 0 | 0 | 255 | blue |
| 0 | 128 | 0 | green | 128 | 0 | 255 | violet |
| 0 | 128 | 128 | teal | 255 | 0 | 255 | magenta |
| 0 | 0 | 128 | navy | 255 | 0 | 128 | rose |
| 128 | 0 | 128 | purple | 255 | 128 | 128 | salmon |
| 128 | 128 | 128 | gray | 255 | 255 | 128 | maize |
| 255 | 0 | 0 | red | 128 | 255 | 128 | mint |
| 255 | 128 | 0 | orange | 128 | 255 | 255 | aqua |
| 255 | 255 | 0 | yellow | 128 | 128 | 255 | periwinkle |
| 128 | 255 | 0 | chartreuse | 255 | 128 | 255 | pink |
| 0 | 255 | 0 | lime | 255 | 255 | 255 | white |
| 0 | 255 | 128 | spring | | | | |

## 4  ENGINEERING STUDY: MATERIALS AND APPROACH

We overlaid 27 virtual colors presented via the oHMD onto eleven representative real-world background conditions (illuminated using 95CRI lights as described in section 3). We measured light exiting the oHMD monocle under the three experimental lighting conditions: backgrounds illuminated (no AR oHMD light), oHMD illuminated (no illuminated backgrounds), and both background and oHMD illuminated.

Five of the eleven backgrounds were actual physical materials (Figure 3) representative of outdoor real-world backgrounds: *brick*, *brown foliage*, *green foliage*, *pavement*, and *sand*. We set these real backgrounds in 16" x 24" frames constructed using 2x4 lumber. This allowed us to easily handle the backgrounds and swap them in and out of the testbed without compromising background surfaces. The specific way in which we assembled and framed each background was slightly different depending upon the nature of the physical material. The *brick* background was assembled by stacking several ten-hole perforated bricks into a frame in a fashion consistent with a brick building facade. Both *brown foliage* and *green foliage* backgrounds were created by scavenging leaves from ground litter (mostly fallen oak leaves) and a living plant (*Spathiphyllum cochlearispathum*, or peace lily) respectively; then affixed and pressed to a 1/2" plywood panel using industrial strength spray adhesive. We obtained a slab of *pavement* from a road construction site, and then cut the slab to fit tightly in a frame. As with the foliage backgrounds, we created the *sand* background by adhering new playground sand to a 1/2" plywood panel. We mounted each of the three panels into a frame using four set screws.

Four of the backgrounds were painted posters (Figure 3) also representing the color of typical real-world backgrounds: *brick poster*, *green foliage poster*, *pavement poster* and *sidewalk poster*. By using these posters, we were able to make interesting comparisons between poster-based backgrounds and backgrounds made of physical materials. Note that there was no attempt to color-match poster backgrounds with their physical material (real) counterparts; although in two cases the matching was very close (*brick* and *pavement*).

We used two additional backgrounds: a *no-lights* condition where the fluorescent lights were turned off, and a *white* condition where we placed a white poster in our apparatus.

We derived 27 colors (presented via the oHMD) to overlay onto backgrounds using combinations of fully saturated (256), half saturated (128) and desaturated (0) values for red, green, and blue (Figure 4). We turned the oHMD off for the black (0,0,0) color condition, in essence measuring the background color as seen through the oHMD's display optics. To automate the projection of color via the oHMD, we created a PowerPoint presentation containing 26 full-screen colored slides. Each colored slide was rendered to the oHMD for two minutes as we continuously recorded data from the colorimeter. On average, we collected approximately two readings per second, although the rate varied according to the color being measured. As a result, we collected between 1459 and 7595 readings per background, for a total of $N$ = 31,195 data points across eleven backgrounds.

We next investigated the colorimeter's behavior over time by studying time-series graphs of each background by color combination. Based on these graphs, we calculated each background by color cell by taking the median $xyY$ value (the colorimeter reports color using the CIE 1931 $xyY$ format, see [16] for details). These medians reduced our collected dataset to one value per background by color combination, giving $N$ = 296 data points[1].

Lastly, we transformed the data into the CIE 1976 $u'v'$ and $L^*u^*v^*$ color spaces, using the formulas found in Foley et al. [17]. There is a direct linear relationship between $xyY$ and $u'v'$, but converting from $xyY$ to $L^*u^*v^*$ requires that the $xyY$ colors be normalized relative to a *white point*. For our experimental setup, this white point represents the brightest possible color (the color with the largest energy density). Our experimental setup yielded three different white points: (1) only the oHMD illuminated (the no-lights condition) and the oHMD presenting the color white (255,255,255), (2) the oHMD off (the color black) against the illuminated white background, and (3) the oHMD presenting the color white against the illuminated white background. Our collected data contained $xyY$ values for each of these conditions; we used these values for our three white points. We used white point (1) when only the oHMD was illuminated, white point (2) when only the background was illuminated, and white point (3) when both were illuminated; we only used these white points to covert $xyY$ values into $L^*u^*v^*$ values. Our calculations used the following formulas (Foley et al. [18]):

$$X = \frac{x}{y}Y, \quad Z = \frac{1-x-y}{y}Y; \quad (1)$$

$$u' = \frac{4X}{X + 15Y + 3Z}, \quad v' = \frac{9Y}{X + 15Y + 3Z}; \quad (2)$$

$$u'_n = \frac{4X_n}{X_n + 15Y_n + 3Z_n}, \quad v'_n = \frac{9Y_n}{X_n + 15Y_n + 3Z_n}; \quad (3)$$

$$L^* = 116(Y/Y_n)^{1/3} - 16, \quad Y/Y_n > 0.01, \quad (4)$$

$$u^* = 13L^*(u' - u'_n), \quad v^* = 13L^*(v' - v'_n);$$

where (eq 1) transforms $xyY$ into 1931 $XYZ$ CIE primaries, (eq 2) transforms $XYZ$ into 1976 CIE $u'v'$ values, (eq 3) transforms $X_nY_nZ_n$, the color of the relevant white point, into corre-sponding $u'_nv'_n$ values, and (eq 4) transforms $Y$, $Y_n$, $u'v'$, and $u'_nv'_n$ into $L^*u^*v^*$ values.

---

[1] There were 10 (*background*: brick poster, real brick, real brown foliage, green foliage poster, real green foliage, pavement poster, real pavement, real sand, sidewalk poster, white poster) × 27 (color) + 1 (no lights) × 26 (all colors except black) = 296 experimental cells; the *no-lights*, *black* condition yields no light energy and was not included.

## 5 RESULTS AND DISCUSSION

To analyze the data, we created a small-multiples graph that depicted all 55 pairwise combinations of our 11 backgrounds. Our motivation for looking at pairwise combinations is the idea that an AR user would be facing one background, and then turn to face another background; we wanted to see the effect of this on the displayed colors. We then printed this large graph on a color printer, cut out each small graph, and then hand-sorted the graphs into various groups and orderings, looking for patterns in how the colors changed against different backgrounds.

Figure 5 shows the result of this analysis. Each small graph depicts a scatterplot in $u'v'$ chromaticity space, and lists a pair of backgrounds in the title strip. On the scatterplot we drew a vector for each color; for a given vector one endpoint (colored white) is the $u'v'$ value of the color measured against the first background, and the other endpoint (colored gray) is the $u'v'$ value measured against the second background. We did not label each color, but instead examined the general pattern formed by the lines. For example, the upper-left small graph of Figure 5 compares the no-lights and the white poster backgrounds. In this graph, we see a star-shaped pattern: against the white poster, the colors are clustered in a small region around the white point, and against the no-lights background the colors are spread throughout $u'v'$ space.

Each small graph also contains a stacked bar graph, which depicts the average change in CIE 1976 $L^*u^*v^*$ space when going from one background to the other. Because $L^*u^*v^*$ space is an approximately perceptually linear 3D space, we calculated 3D Euclidean norm distances in this space. For each of our 27 colors, we calculated the 3D Euclidean distance for each pair of backgrounds (for backgrounds paired with the *no-lights* condition we calculated 26 distances). Each of these distances is the length of a vector in $L^*u^*v^*$ space, and represents a total perceptual color change. We can consider this vector to consist of two components: $u^*v^*$, which represents a change in chromaticity, and $L^*$, which represents a change in luminance. In order to consider chromaticity and luminance separately, we calculated the length of $L^*$ projected onto the vector $L^*u^*v^*$; this length gives the proportion of the total perceptual color change that is due to luminance. This gave us 27 (26) pairs of values per background pair.

We next examined the distribution of these values over the background pairs, and determined that the average was a reasonable measure of central tendency.

In Figure 5, the stacked bar graphs depict these calculations. Together, the pink and blue bars depict the total average length of $L^*u^*v^*$ for each background pair. The pink bar represents the average projection of $L^*$ on $L^*u^*v^*$. Our interpretation is that the pink bar measures luminance change, the blue bar measures chromaticity change, and both bars together measure total perceptual color change.

The upper-left graph, which compares the no-lights and the white poster backgrounds, has the largest perceptual color change; the left-hand edge of this bar has the value 0, and the right-hand edge has the value 116.6, in $L^*u^*v^*$ units. All of the other bar graphs are scaled relative to this maximum value.

After printing and cutting out these 55 small multiples, we considered the information contained in both the $u'v'$ scatterplot and the $L^*u^*v^*$ stacked bar chart. Figure 5 shows our final arrangement: rows are ordered according to the first background in each pair, while columns are ordered according to the second background in each pair. The colored borders indicate the following five categories:

**Washout Due to Chromaticity (a)**: This category consists of the 10 background pairs in the top row of Figure 5, where the no-lights background is paired with all of the other backgrounds. Figure 6a shows a larger version of the $u'v'$ scatterplot of one of these pairs, the white poster / no-lights combination; these larger scatterplots allow an analysis of how each oHMD color changes when measured against these two backgrounds. In this category, in the no-lights condition, the colors are widely distributed in $u'v'$ space. This is equivalent to using an AR system on a dark night, where we would expect a large hue separation between the display colors. When measured against an illuminated background, the colors collapse into a much smaller point, clustering around the background color, and creating a large star-shaped pattern. This causes the hue differences to appear washed out — a common AR experience outdoors. For example, in Figure 6a the rightmost data point (a black diamond (♦)) represents the color red in the no-lights condition; under the white poster background this color has shifted leftwards and is part of the central grouping of colors. This indicates that the color red is quite vibrant and saturated in the no-lights condition, but becomes desaturated against white, and may even no longer appear to be red (although verifying this would require a human color judgment).

The unexpected element of this analysis comes from studying the bar graphs for this group: these indicate that the great majority of the perceptual change is due to a large chromaticity shift for each color. Surprisingly, the bar graphs indicate that the change in luminance is a relatively minor perceptual component when going from complete darkness to an illuminated background.

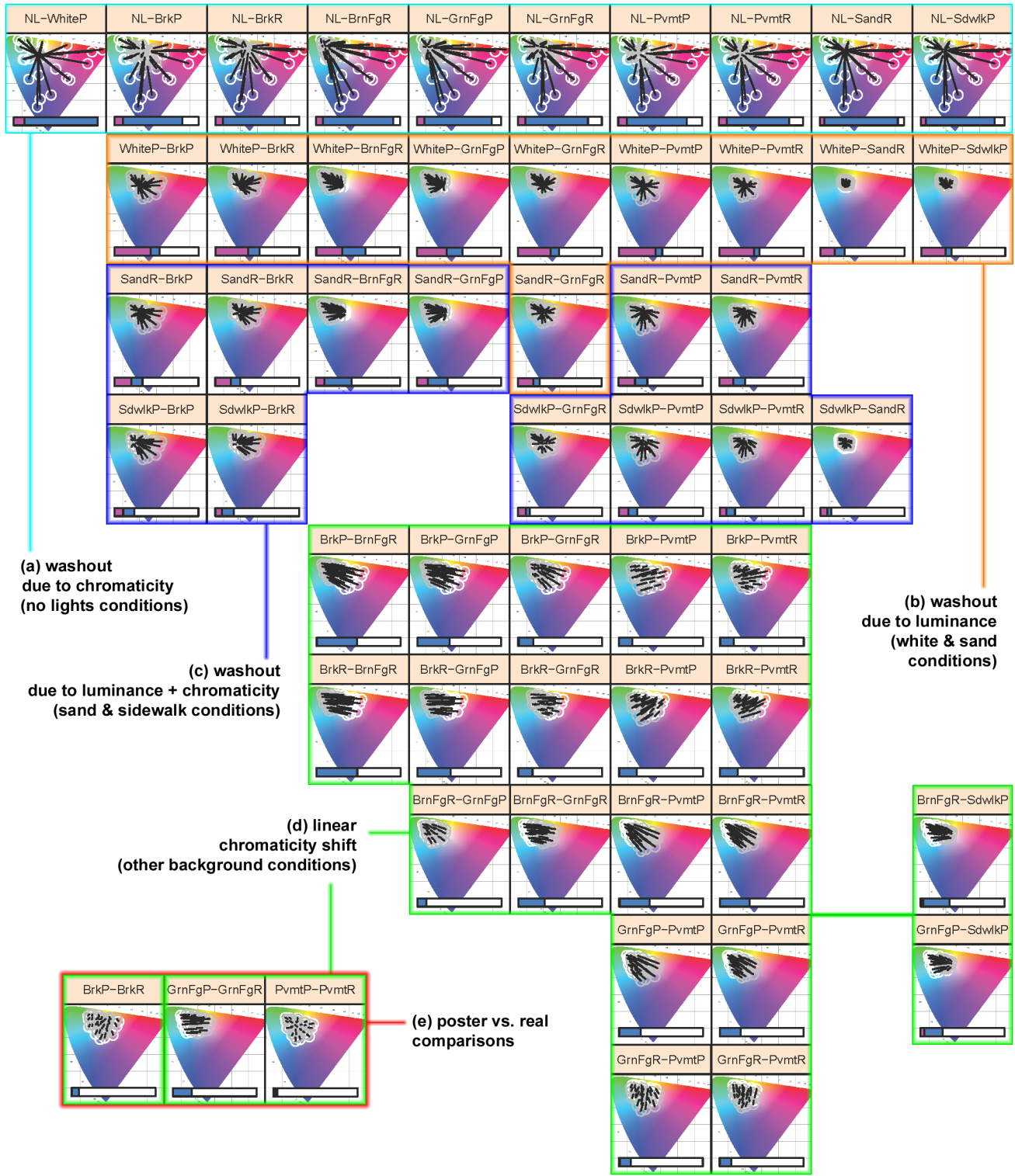

**Figure 5**. Pairwise visual analysis of eleven backgrounds. Small multiples of scatterplots are outlined into four categories (a–d), depending upon the nature of color shifts: (a) washout due to chromaticity, (b) washout mostly due to luminance, (c) washout due to both chromaticity and luminance, and (d) linear shift in chromaticity. Category (e) compares the three poster-real conditions. See the text for a detailed description of the small multiple scatterplots. From top to bottom, scatterplots are grouped according to the first background denoted in the pair. From left to right, scatterplots are ordered alphabetically according to the second background in the pair. Each scatterplot is labeled to denote the pair of backgrounds depicted using NL (*no light*), White (*white*), Brk (*brick*), BrnFg (*brown foliage*), GrnFg (*green foliage*), Pvmt (*pavement*), Sand (*sand*), and Sdwlk (*sidewalk*). Labels ending in an 'R' denote a real physical material background; labels ending in a 'P' denote a poster background.

This analysis raises two potential concerns: (1) will UI components become less legible against the bright white background? And, (2) will the hues in the center become less perceptually discriminable, meaning that fewer hues are available as a UI design component? Figures 5 and 6 suggest that both effects may occur, but human judgments are required to validate both concerns.

**Washout Due to Luminance (b)**: This category consists of the 10 background pairs in the second row of Figure 5, where the white poster is paired with every other background, as well as the sand / real green foliage combination from the third row. Here the washout effect is broadly similar to (a), but the chromaticity change is greatly reduced, appearing as a much smaller star-shaped pattern. Figure 6b shows a representative example from this group, the white poster / real green foliage combination. For this group both backgrounds are relatively bright, and so the colors remain washed out; the bar graphs indicate that the majority of the perceptual color change is due to a change in luminance.

**Washout Due to Luminance + Chromaticity (c)**: This category consists of 12 background pairs in the third and forth rows of Figure 5, where the sand and sidewalk poster backgrounds are paired with other backgrounds. Figure 6c shows a representative example (real sand / real pavement). The pairs in this category are broadly similar in appearance to (b), in that they all show a similar-sized washout effect and star-shaped pattern. However, here the bar graphs indicate that luminance and chromaticity contribute relatively evenly to the overall perceptual color change. We note that there is not a clear separation between this category and (b), and several background pairs placed here could also be placed in (b).

**Linear Chromaticity Shift (d)**: This category consists of the remaining 23 background pairs, where the less bright backgrounds are compared to each other; Figure 6d shows a representative example (real brown foliage / real brick). Here the shape in the $u'v'$ scatterplots indicates a linear shift in all of the colors; for example, in Figure 6d, when the user moves from a brown foliage background to red bricks, all of the oHMD colors are shifted from left to right, towards the red side of the chromaticity diagram. The bar graphs indicate that luminance contributes very little; almost all of the perceptual color change is due to this linear shift in chromaticity. The primary usability concern from such a linear shift is situations where hues change; for example in Figure 6d yellows shift into oranges, greens shift into yellows, and blues shift into purples. While Figure 6d suggests that there are a variety of hues available for UI components against both backgrounds, the shifting of hues could be problematic in domains where color encoding is critically important.

**Poser vs. Real Comparisons (e)**: This category is a selection of three background pairs from category (d), but here the interest is that the real backgrounds of brick, green foliage, and pavement are each compared with their respective poster versions. In Figure 5(e) we compare the posters and the real backgrounds; Figure 7 shows a larger version of these scatterplots. As with the rest of category (d), there was a linear color shift for each pair. However, for the real / poster pavement condition (Figure 7c) we measured the smallest $L^*u^*v^*$ color change in the entire experiment (5.6 $L^*u^*v^*$ units). Figure 7a, comparing the brick poster with a background of real bricks, shows a small star-shaped pattern, while Figure 7b, comparing the green foliage poster with a background of actual green leaves, shows a small linear shift.

A finding of this study is that the patterns and trends described above related to physical materials as real-world backgrounds are broadly similar to those patterns reported in other related studies that employed color-matched, painted posters as real-world backgrounds (e.g., [1]). We were somewhat surprised that there were not larger color changes between the real and poster versions of these backgrounds, especially considering that the painted posters were only a metameric color match against real-world backgrounds.

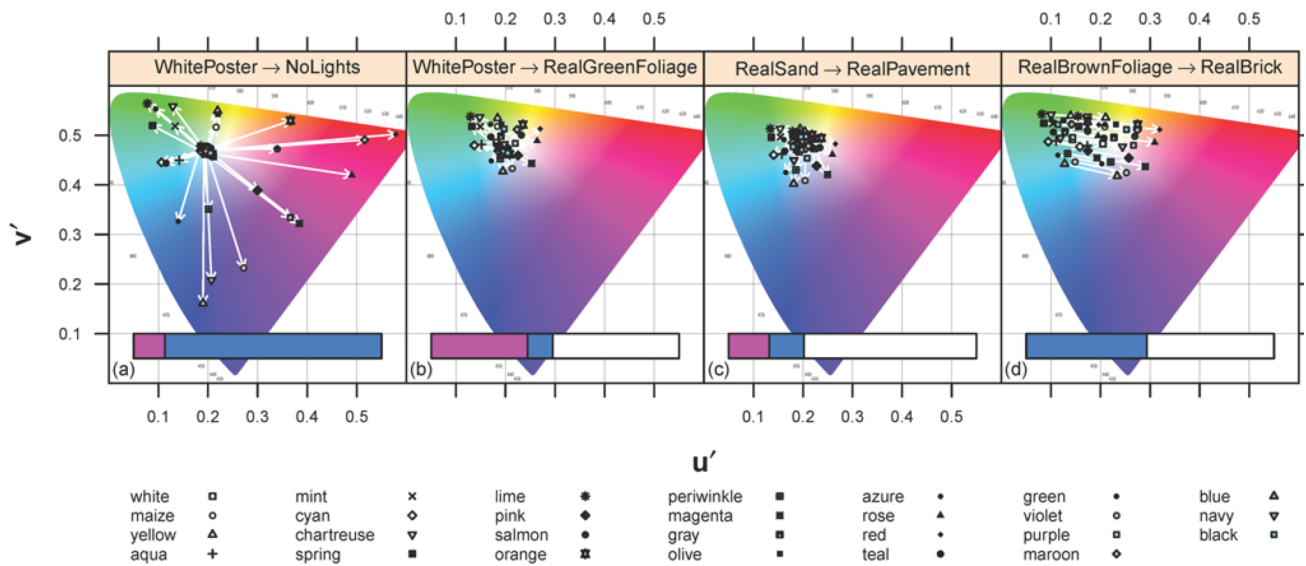

**Figure 6**. Pairwise visual analysis of four background pairs. Each panel is an example from the first four categories depicted in Figure 5: (a) washout due to chromaticity (white poster → no lights); (b) washout mostly due to luminance (white poster → real green foliage); (c) washout due to both chromaticity and luminance (real sand → real pavement); and (d) linear shift in chromaticity (real brown foliage → real brick). Each panel shows how every color changes when the background changes; for each color, the arrow points from the first background towards the second background as indicated in the panel title. The cyan square is for the color black, when the HMD is turned off; this is the color of each background as measured through the HMD's optics. Note that in (a) there is only one cyan square; this is the poster + lights white point.

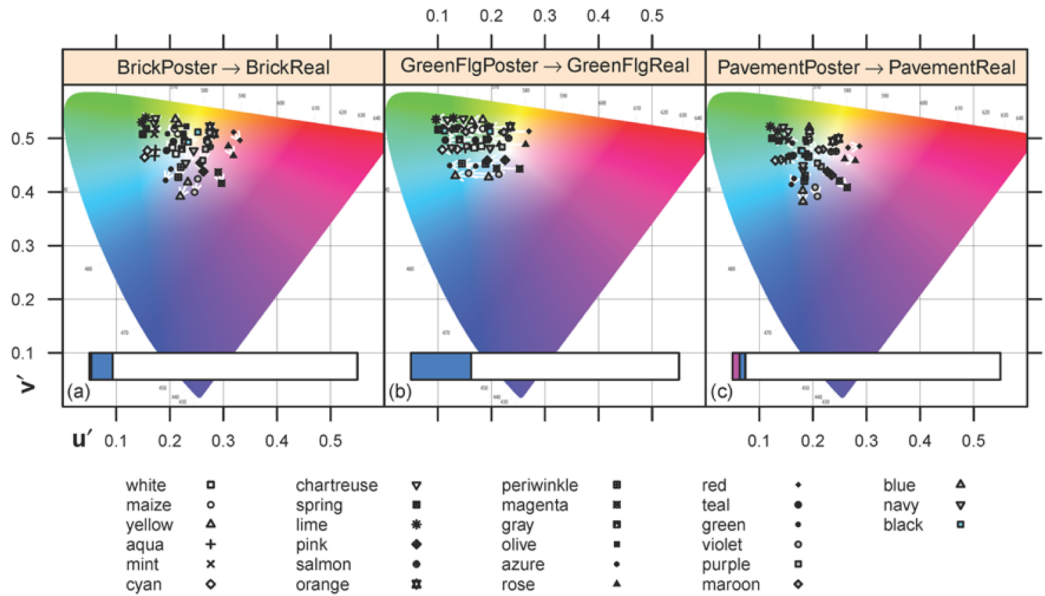

This finding is one argument for the utility of color-matched posters for this type of analysis. Indeed, since posters can be created more readily and are easier to handle, these types of engineering studies can be used to examine a large set of independent variables (e.g., backgrounds, AR colors, source lighting) in a relatively short amount of time. However, the texture, depth, and reflectance properties of painted posters may not be comparable to their respective physical material.

Lastly, since this work is an engineering study that measured quantifiable elements, we need to conduct follow-on user studies to verify the perceptual consequences we conjecture above. From our own experiences (and others), we know that washout is common and is obvious to most causal outdoor oHMD users. However, we have not seen a discussion in the literature related to linear chromaticity shift. One explanation may be color constancy; a feature of the human perception system that ensures an object's perceived color remains constant under illumination changes that alter the object's measured color [19].

## 6 Future Work

In a future study, we will consider independently varying the luminance of both the lighting source and oHMD. Using this study as a rich set of baseline data, we will then be able examine how differing levels of luminance contrast (between light source and oHMD) affects color shifts in AR graphics against various real-world backgrounds. Of specific interest would be integrating an oHMD with much higher luminance to examine the degree to which simply building brighter displays mitigates these color blending effects. Based on our work to date, we are confident that there are important chromaticity interactions to consider independent of luminance.

We plan to conduct a user-based perceptual color-matching study using these same independent variables to compare how users' perception matches these data. If the data matches under some conditions, then we can subsequently collect volumes of engineering data with some assurance that resulting analyses will provide insight into actual user experiences.

We are planning a mobile version of the testbed to use both indoors and outdoors in arbitrary locations where varied, dynamic and/or combined white points are present to better understand how colors perception is effected when users move about.

This planned work, in conjunction with other work in the field, can be used to inform a predictive model of color blending. Longitudinally, we hope to apply this predictive model to adaptive AR systems that make real-time adjustments (based on the environment) to the UI color presentation layer to preserve color encoding and designer intent.